\documentclass[12pt]{iopart}
\usepackage{iopams,epsf}

\usepackage{color}
\usepackage{epsfig}
\usepackage{graphics}
\usepackage{psfrag}
\usepackage{lscape}

\newcommand{\sts}{\scriptsize}

\newcommand{\mb}{\mbox}
\newcommand{\p}{\partial}

\begin{document}

\title[Spherical Model in the microcanonical formalism]{Critical properties of the spherical model in the microcanonical formalism}

\author{Hans Behringer}

\address{Fakult\"at f\"ur Physik, Universit\"at Bielefeld,
  Universit\"atsstra\ss e 25, D -- 33615 Bielefeld, Germany}

\ead{Hans.Behringer@physik.uni-bielefeld.de }

\begin{abstract}
Due to the equivalence of the statistical ensembles
thermostatic properties of physical systems with short-range
interactions can be calculated in
different ensembles leading to the same physics. In particular,
the ensemble equivalence holds for systems that undergo a
continuous phase transition in the infinite volume limit so that the
properties of the transition can also be investigated in the
microcanonical approach. Considering as example the spherical
model the ensemble equivalence is explicitly demonstrated by
calculating the critical properties in the microcanonical
ensemble and comparing them to the well-known canonical results.

\end{abstract}




\section{Introduction}

The properties of physical systems exhibiting a continuous (or
discontinuous) phase transition are usually investigated in the
canonical approach. However, for systems with short-range
interactions  the various physical ensembles such as the canonical
ensemble and the microcanonical one are equivalent in the infinite
volume limit so that the properties of continuous phase transitions can be
investigated in the microcanonical ensemble as well.\footnote{It should be
noted that the ensembles are inequivalent at a first order transition
point even if the interactions are short-ranged. This subtle
difference is exploited for the study of phase separation in
microcanonical systems (e.\,g. \cite{Gross_Votyakov_2000}).} The microcanonical
description is based on the entropy as the thermodynamic potential
and  the physical properties of the system are deduced from its
geometry. Microcanonical response functions, for example, are
related to the curvature of the entropy surface.

The description of phase transitions in the microcanonical formalism has
gained growing interest in recent years (\cite{Gross_1990, Hueller_1994,
Promberger_Hueller_1995, Gross_etal_1996, Gross_2001} and
references therein). Apart
from works about discontinuous phase transitions  in
microcanonical systems \cite{Brown_Yegulalp_1991, Schmidt_1994, Wales_Berry_1994, Gross_etal_1996,
Deserno_1997, Ota_Ota_2000, Gross_2001, Ispolatov_Cohen_2001}  second order
phase transitions have been studied recently
\cite{Promberger_Hueller_1995, Bruce_Wilding_1999, 
Kastner_etal_2000, Hueller_Pleimling_2002, Behringer_2003, Behringer_2004, Hove_2004, Naudts_2004}. Ways to extract critical
exponents from microcanonical quantities calculated for finite
systems have also been suggested and applied successfully for various
model systems \cite{Kastner_etal_2000,
Hueller_Pleimling_2002, Pleimling_etal_2004, Behringer_etal_2005}. All these works basically concentrate
on signatures of phase transitions in finite systems although some
works investigated the general scaling behaviour of the entropy of the
infinite system near a continuous transition \cite{Kastner_etal_2000,
Behringer_2005}.

The main purpose of this paper is to demonstrate the equivalence
of the microcanonical and the canonical ensemble in the infinite
system for a concrete example of a model system with short-range
interactions that undergoes a
second order phase transition and can be
tackled analytically.\footnote{See
for instance \cite{Dauxois_etal_2000, Barre_etal_2001,
Ispolatov_Cohen_2001, Kastner_2005} for recent investigations of the
question of the equivalence of the microcanonical and the canonical ensemble
for systems with long-range forces.} To this end the  properties of the spherical
model at the phase transition point are investigated within the
microcanonical ensemble. By doing so it is also shown exemplarily
that the investigation of continuous phase
transitions is not restricted to the canonical ensemble and can be
carried out auspiciously in the microcanonical ensemble, too. There
are only a few model systems with short-range interactions 
and a phase transition in the thermodynamic limit which, within 
the framework of the canonical ensemble, have been solved exactly.
The zero-field Ising model in two dimensions and the spherical model in all 
dimensions belong to the class of models where the free energy 
density of the infinite system is known analytically. For all the
author knows a microcanonical investigation of a model system with
short-range forces exhibiting a non-trivial continuous phase
transition in the thermodynamic limit has not been reported in the
literature yet. Here an
investigation of the interrelation of the canonical and
microcanonical critical exponents of the spherical model is
carried out. As the natural variables are different for the
microcanonical and the canonical ensemble the critical exponents
are in general not identical. The equivalence of the ensembles,
however, leads to a relation between them. Note that this relation
was discussed in \cite{Hankey_Stanley_1972} for statistical ensembles
whose thermodynamic potentials are connected to each other by
Legendre transforms. The entropy as the thermodynamic potential of
the microcanonical ensemble, however, is related, for instance, to the free
energy by a Legendre transform and a subsequent
partial inversion. The results for the microcanonical spherical
model presented in this paper corroborate general considerations
that have been based on scaling relations for the microcanonical
entropy function \cite{Kastner_etal_2000, Behringer_2005}.

The rest of the paper is organised as follows. In section
\ref{kap:thermo} a brief introduction to the microcanonical
analysis of physical properties of ferromagnetic systems is
given. This section also sets up the notation and language used later on.
The specific entropy of the ferromagnetic spherical model is
calculated in section \ref{kap:spezentro} in the macroscopic limit
using the method of steepest descent. The critical properties of
the spherical model are then analysed microcanonically in section
\ref{kap:kriteigen} with special focus laid on the values of the
microcanonical critical exponents. Section \ref{kap:mean} contains
some comments on the mean spherical model. The finding are summarised and
compared to the canonical results in the final section
\ref{kap:zus}.

\section{Thermostatics in the microcanonical formalism}
\label{kap:thermo}

The basic quantity in the investigation of the statistical
properties of a finite magnetic system with $N$ particles is the density of
states
\begin{equation}
\label{gl:defdossallg} \Omega_N(E,M) = \int \mbox{d}\Gamma_N
\delta(E-\mathcal{H}(\sigma))\delta(M-\mathcal{M}(\sigma)).
\end{equation}
The Hamiltonian $\mathcal{H}$ provides  the energy for any
microstate $\sigma= (\sigma_1,\ldots,
\sigma_N)$ from the phase space $\Gamma_N$ of all
possible configurations of the $N$-particle system. The
magnetisation $M$ of the configuration $\sigma$ is measured by the operator $\mathcal{M}$.
Usually, the density of states is Laplace-transformed to yield the
canonical partition function which determines the free
energy. Note that the free energy is the thermodynamic potential of the canonical ensemble. In this
paper, however, the density of states or equivalently the
microcanonical entropy is directly analysed to deduce
the physical properties of the spherical model.

The microcanonical entropy density of a finite magnetic system with $N$
particles is obtained from the density of states by taking the
logarithm
\begin{equation}
  s_N(e,m) = \frac{1}{N} \ln\Omega_N(eN, mN)
\end{equation}
where the energy density is defined by $e=E/N$, analogously, the
magnetisation density is given by $m = M/N$. Note that units with
$k_{\sts \mb{B}} = 1$ are used in this work. The thermodynamic properties of the system in the thermodynamic limit
are calculated from the entropy
\begin{equation}
\label{gl:defspezentropie}
  s(e,m) = \lim_{N\to \infty} s_N(e,m).
\end{equation}

The thermostatics of an infinite system can be investigated
by studying the free energy as the thermodynamic potential with the
temperature being one of the natural variables or alternatively by
considering the entropy where the energy shows up as a natural variable \cite{Callan_1960}. In the following it is
briefly summarised how the physical properties of statistical systems can
be deduced from the entropy function $s(e,m)$. The inverse temperature
$\beta(e,m)$ and the magnetic field $h(e,m)$ are basically given
by the first derivatives of the entropy function:
\begin{equation}
\beta(e,m) = \frac{\p }{\p e} s(e,m)
\end{equation}
and
\begin{equation}
h(e,m) = - (\beta(e,m))^{-1} \frac{\p}{\p m} s(e,m).
\end{equation}
Note that the inverse temperature $\beta$ and the magnetic field $h$ show up
as conjugate variables to the natural variables $e$ and $m$ in the
microcanonical ensemble.

The zero-field macrostate $(e,m_{\sts \mb{sp}}(e))$ is defined to
be the state with zero magnetic field $h(e,m_{\sts \mb{sp}}(e)) =
0$ for given energy $e$. The associated magnetisation $m_{\sts
\mb{sp}}(e)$ is called spontaneous magnetisation. The
corresponding inverse temperature is defined to be $\beta_0(e) =
\beta(e,m_{\sts \mb{sp}}(e))$. The response functions of the
system are related to second order derivatives of the entropy function
\cite{Kastner_etal_2000, Behringer_2004, Behringer_etal_2005}. The
specific heat for the  zero-field macrostate, for instance, is explicitly
given by
\begin{equation}
\label{def:spez} c_0(e) = -\frac{(\beta_0(e))^2}{\frac{\p}{\p
e}\beta_0(e)}.
\end{equation}

In the vicinity of a critical point, which shows up at a
critical energy $e_{\sts \mb{c}}$, the physical quantities display
power-law behaviour. Denoting the deviation of the energy $e$ from
the critical value $e_{\sts \mb{c}}$ by $\varepsilon := e -
e_{\sts \mb{c}}$ a general physical quantity $a$ has the form
$a(e) \sim |\varepsilon|^{-\varkappa_\varepsilon}$ for
small $\varepsilon$. The singularity of $a$ is
characterised by the critical exponent $\varkappa_\varepsilon$ in
the microcanonical formalism. The critical exponent $\varkappa_t$ which
characterises the singularity of $a$ in the canonical
formalism
(i.\,e. $a(T) \sim |T-T_{\sts \mb{c}}|^{-\varkappa_t}$ with $T$
being the temperature and $T_{\sts \mb{c}}$ being the critical
temperature) is related to the exponent $\varkappa_\varepsilon$ by
the equation
\begin{equation}
\label{gl:exporel} \varkappa_\varepsilon = \frac{\varkappa_t}{1 -
\alpha_t}.
\end{equation}
The critical exponent of the specific heat in the canonical
formalism is denoted by $\alpha_t$ and has to be in the interval
$]0,1[$ for relation (\ref{gl:exporel}) to be valid \cite{Promberger_Hueller_1995}. If the
specific heat in the canonical formalism has a jump singularity,
a logarithmic singularity or a cusp singularity at the transition
point the microcanonical and canonical critical exponents are
identical.

As this paper considers the spherical model, which has
a cusp singularity in the specific heat, the case of a cusp
singularity is briefly sketched. In the vicinity of the
transition point the canonical specific heat with a cusp at the
transition temperature $T_{\sts \mb{c}}$ has the general form
\begin{equation}
\label{gl:negspezcan} c^{(\sts \mb{c})} \sim A + B_{\pm}
|t|^{-\alpha_t}
\end{equation}
with $t:= T - T_{\sts \mb{c}}$ and a negative canonical
exponent $\alpha_t$. Integrating expression (\ref{gl:negspezcan})
gives
\begin{equation}
    \varepsilon \sim At - \frac{B_\pm}{1+|\alpha_t|}
    |t|^{1+|\alpha_t|}.
\end{equation}
The dominating term in the limit $t\to 0$ is thus the linear
term and one has $\varepsilon \sim t$ near the transition point.
Therefore, physical quantities have the same qualitative
dependence when expressed as functions of the reduced temperature
$t$ or the reduced energy $\varepsilon$. The critical exponents
are consequently identical for the canonical and microcanonical
description.

The exponent $\delta_\varepsilon$ describes the relation between
the critical magnetic field and the magnetisation at the
transition energy (or temperature). The microcanonical exponent is
therefore always identical to the canonical exponent $\delta_t$.

\section{Specific entropy of the spherical model}
\label{kap:spezentro}

In this section the density of states of the spherical model
\cite{Berlin_Kac_1952} is calculated for finite
systems. The spherical model exhibits a continuous phase
transition and can be solved analytically for an arbitrary
magnetic field in any dimension $d$. Therefore, its critical
properties have been studied intensely within the canonical
ensemble in the past (see e.\,g. \cite{Joyce_1972, Baxter_1982, Thompson_1988}).  
From the density of states one gets the specific entropy
of the infinite lattice by carrying out the thermodynamic limit
(\ref{gl:defspezentropie}). The density of states will be
evaluated for a hyper-cubic system in $d$ dimensions with a linear
extension $L$ so that the system contains $N= L^d$ spins. The spin
variables $\sigma_i \in \mathbb{R}$, $i = 1, \ldots, N$, can take
on any real value, but they have to satisfy the constraint
\begin{equation}
\label{gl:nebenbedsph}
  \sum_{i=1}^N \sigma_i^2 = N.
\end{equation}
The phase space of the spherical model is therefore the sphere of
radius $\sqrt{N}$ in $\mathbb{R}^N$. The integration measure in the
definition (\ref{gl:defdossallg}) of the density of states is just
given by the Lebesgue measure $\mbox{d}\Gamma_N = \mbox{d}^N
\sigma$.
The Hamiltonian of the spherical model is given by
\begin{equation}
\label{gl:gausshamil}
    \mathcal{H}(\sigma) = - J \sum_{\left< i,j \right
    >} \sigma_i \sigma_j
\end{equation}
with a positive exchange constant $J$, the magnetisation is 
\begin{equation}
    \mathcal{M}(\sigma) = \sum_i \sigma_i.
\end{equation}
The angular brackets $\left < i,j\right >$ indicate a summation over
all neighbouring lattice sites $i$ and $j$. The Hamiltonian
(\ref{gl:gausshamil})  together with the subsidiary condition
(\ref{gl:nebenbedsph}) defines the ferromagnetic spherical model with nearest neighbour
interactions only.

The density of states of the spherical model is generally given by
\begin{equation}
\label{gl:entro_deltarestriction}
    \Omega_N(E,M) = \int\limits_{\mathbb{R}^N}
    \mb{d}^N\sigma \, \delta(E -
    \mathcal{H}(\sigma))\delta (M - \mathcal{M}(\sigma))\delta(N-\sum_i\sigma_i^2).
\end{equation}
In view of the three delta functions the integral remains unchanged
if one inserts the factors $\exp\left(a E-a\mathcal{H}(\sigma)\right)$,
$\exp(bM - b\mathcal{M}(\sigma))$ and $\exp(cN-c\sum_i\sigma_i^2)$ with real $a$, $b$ and $c$. Using the
Fourier representation
\begin{equation}
    \delta(x) =  \int \frac{ \mb{d}k}{2\pi}\exp(ikx)
\end{equation}
of the delta function one can rewrite the expression of the
density of states as
\begin{equation}
\Omega_N(E,M) = \int \frac{ \mb{d}p}{2\pi}\int \frac{
\mb{d}q}{2\pi} \int \frac{ \mb{d}r}{2\pi} \int
    {\mb{d}^N\sigma} \exp \left(
    A_{p,q,r}(\sigma)\right),
\end{equation}
where the argument $A$ of the exponential is given by
\begin{equation}
A_{p,q,r}(\sigma) = (a+ip)E + (b+iq)M + (c+ir)N + Q_{p,q,r}(\sigma)
\end{equation}
with the quadratic form
\begin{eqnarray}
\label{gl:defwmatrix} 
Q_{p,q,r}(\sigma) &=& (a+ip) \sum_{\left<
i,j\right>} \sigma_i\sigma_j - (c+ir)\sum_i \sigma_i^2 -
(b+iq)\sum_{i}\sigma_i \nonumber\\ 
&=& - \sigma^{\mb{\sts T}}W\sigma + v^{\mb{\sts T}}\sigma.
\end{eqnarray}
The last equality defines the matrix $W$, which describes the
interaction of the spins, the vector $v$ is the vector in $\mathbb{R}^N$ which has
$N$ identical entries $-(b + iq)$. The transposed vector of $v\in
\mathbb{R}^N$ is denoted by $v^{\mb{\sts T}}$. Introducing new
spin variables $\mu \in \mathbb{R}^N$ by
\begin{equation}
    \mu = \sigma - \frac{1}{2} W^{-1} v
\end{equation}
the form $Q$ is given by
\begin{equation}
    Q_{p,q,r}(\mu) = - \mu^{\mb{\sts T}} W \mu + \frac{1}{4}v^{\mb{\sts T}} W^{-1} v
\end{equation}
in terms of the new variables $\mu$ (provided the inverse $W^{-1}$
exists). The spin variables $\mu$ appear now quadratically 
so that one is left with a multiple Gaussian integral yielding
\begin{equation}
\Omega_N(E,M) = \int \frac{ \mb{d}p}{2\pi}\int \frac{
\mb{d}q}{2\pi} \int \frac{ \mb{d}r}{2\pi}  \exp\left(
\Phi_{p,q,r}(E,M)\right)
\end{equation}
with the argument
\begin{equation}
\Phi_{p,q,r}(E,M) =  (c+ir)N + (a+ip)E + (b+iq)M + \frac{1}{2}v^{\mb{\sts T}} W^{-1} v -
\frac{1}{2} \ln \det W.
\end{equation}
Here the identity $(\det W)^{-1/2} = \exp(- \frac{1}{2} \ln \det
W)$ has been used.

The interaction matrix $W$, that has been defined in expression
(\ref{gl:defwmatrix}), is a generalised cyclic matrix, which can be
diagonalised by a Fourier transformation in any dimension $d$
(e.\,g. \cite{Baxter_1982, Thompson_1988}). For the case of nearest-neighbour
interactions only one gets the eigenvalues
\begin{equation}
    \lambda(\varphi_1, \ldots, \varphi_d) = (c+ir) - (a+ip)J
    \sum_{j=1}^d \cos \varphi_j
\end{equation}
with $\varphi_j = (l-1)2\pi/L$, $l = 1, \ldots, L-1$ and $j = 1,
\ldots , d$. The logarithm of the determinant of $W$ is now given
by
\begin{equation}
    \ln \det W \stackrel{N \gg 1}{\sim} N
    \int\limits_{[0,2\pi]^d}\frac{\mb{d}^d\varphi}{(2\pi)^d} \ln \left((c+ir)- (a+ip)J\sum_{j=1}^d \cos \varphi_j
    \right)
\end{equation}
where the summation has been replaced by an integral in the macroscopic limit of asymptotically large $N$ (i.\,e. large $L$).
As $v$ is an eigenvector of $W$ with the eigenvalue $(c+ir) - (a+ip)dJ$
it is also an eigenvector of
the inverse $W^{-1}$ with the eigenvalue $((c+ir) - (a+ip)dJ)^{-1}$. Therefore, one
has
\begin{equation}
    \frac{1}{4} v^{\mb{\sts T}} W^{-1} v = \frac{N}{4((c+ir)-(a+ip)Jd)} (b+iq)^2.
\end{equation}

Introducing the new integration variables $z = a+ip$, $w = b+iq$
and $u = c+ir$ the density of states is now given by
\begin{equation}
    \Omega_N (Ne,Nm) \stackrel{N \gg 1}{\sim}  \int\limits_{a - i\infty}^{a+i\infty}
    \frac{\mb{d}z}{2\pi} \int\limits_{b - i\infty}^{b+i\infty}
\frac{\mb{d}w}{2\pi}\int\limits_{c - i\infty}^{c+i\infty}
\frac{\mb{d}u}{2\pi} \exp(N\phi_{z,w,u}(e,m))
\end{equation}
with the argument
\begin{eqnarray}
\label{gl:phivorsattelaus}
    \phi_{z,w, u}(e,m) &=& ze + wm + u - \frac{1}{2}
    \int\frac{\mb{d}^d\varphi}{(2\pi)^d} \ln \left(u - zJ\sum_j \cos
\varphi_j\right) \nonumber\\ &&
     + \frac{w^2}{4(u-zdJ)}.
\end{eqnarray}
This expression of the density of states for asymptotically large
$N$ can be used now to calculate the specific entropy of the
infinite system by using the method of steepest descent
\cite{Hilbert_Courant_1953}. The
argument of the logarithm in (\ref{gl:phivorsattelaus}) must have
a strictly positive real part for any possible value of the
$\varphi_j$. Therefore, the real part of $u$ has to be positive. If $\mb{Re\,} z \in~]-\mb{Re\,}u/(dJ),
\mb{Re\,}u/(dJ)[$ the function $\phi$ is analytic in $z$, $w$ and
$u$ in this domain.
Consider the function $\phi$ first for real values of $z$, $w$ and
$u$. For $w\to \pm \infty $ one has $\phi \to \infty $ and for $z
\to \pm u_0/(dJ)$ one also has $\phi \to \infty$. The function
$\phi$ has therefore a minimum for a real $w_0$, $z_0$ and $u_0$
with $z_0 \in~]-u_0/(dJ), u_0/(dJ)[$. Due to the analyticity of
$\phi$ the integration path can now be deformed to have $a=z_0$,
$b=w_0$ and $c = u_0$ so that one obtains
\begin{equation}
\label{gl:gauss-entropie}
s(e,m) = \lim_{N\to \infty} \frac{1}{N} \ln \Omega_N(eN,mN) =
\phi_{z_0,w_0, u_0}(e,m)
\end{equation}
for the specific entropy in the thermodynamic limit by the method of
steepest descent. The values $z_0$, $w_0$ and $u_0$ as  functions of the
energy $e$ and the magnetisation $m$ are determined by the saddle point
equations
\begin{equation}
\label{gl:z-sattelgl}
\frac{\partial \phi}{\partial z} = e + \frac{1}{2} \int
\frac{\mb{d}^d\varphi}{(2 \pi)^d} \frac{J\sum_j \cos \varphi_j}{u - z J
\sum_j \cos \varphi_j} + \frac{dJ}{4(u-zdJ)^2}w^2 = 0,
\end{equation}
\begin{equation}
\label{gl:w-sattelgl}
\frac{\partial \phi}{\partial w} = m + \frac{1}{2(u-zdJ)} w = 0
\end{equation}
and
\begin{equation}
\label{gl:u-sattelgl}
\frac{\partial \phi}{\partial u} = 1 - \frac{1}{2} \int
\frac{\mb{d}^d\varphi}{(2 \pi)^d} \frac{1}{u - z J
\sum_j \cos \varphi_j} - \frac{1}{4(u-zdJ)^2}w^2 = 0.
\end{equation}

\section{Critical properties of the spherical model}
\label{kap:kriteigen}

 In this section the physical properties of the spherical
model are deduced directly from the specific entropy
(\ref{gl:gauss-entropie}). The main focus is laid on the  possible
appearance of a continuous phase transition signalled, for
example, by diverging response functions. In particular, the
character of the singular physical quantities at the phase
transition point which shows up in dimensions larger than two
will be scrutinised. As the entropy of the spherical model is an
even function in $m$ the discussion will consider only
non-negative magnetisations.

\subsection{Discussion of the saddle point equations}

The microcanonical inverse temperature of the spherical model is given
by
\begin{eqnarray}
\beta(e,m) &=& \frac{\partial}{\partial e} s(e,m)= \frac{\p\phi}{\p
e} + \frac{\p\phi}{\p z_0}\frac{\p z_0}{\p e} + \frac{\p\phi}{\p
w_0}\frac{\p w_0}{\p e}  + \frac{\p\phi}{\p u_0}\frac{\p u_0}{\p
e}= \frac{\p\phi}{\p e} \nonumber \\ & =& z_0(e,m)
\end{eqnarray}
where the saddle point equations (\ref{gl:z-sattelgl}),
(\ref{gl:w-sattelgl}) and (\ref{gl:u-sattelgl}) have been used for
the last but one equality. Similarly, the microcanonical magnetic
field is determined by
\begin{equation}
\label{gL.hgleichungmitm} h(e,m)\beta(e,m) = - \frac{\p}{\p m}
s(e,m) = - w_0(e,m).
\end{equation}
Defining the new function $\zeta(e,m) := u_0(e,m)/z_0(e,m)$ the
saddle point equation (\ref{gl:w-sattelgl}) can be re-expressed as
\begin{equation}
\label{gl:w-sattelglneu}
  h(e,m) = 2m (\zeta(e,m) - dJ).
\end{equation}
Note that the variable $\zeta(e,m)$ has to be larger than the
critical value  $\zeta_{\sts \mb{c}} := dJ$ due to the restrictions for
the argument of the logarithm in (\ref{gl:phivorsattelaus}).
The relation (\ref{gl:w-sattelglneu}) can be used now to rewrite
the other two saddle point equations (\ref{gl:z-sattelgl}) and
(\ref{gl:u-sattelgl}):
\begin{eqnarray}
\label{gl:sat1}
    \beta(e,m) \left[ e +dJm^2\right] &=& - \frac{1}{2} \int
\frac{\mb{d}^d\varphi}{(2 \pi)^d} \frac{J\sum_j \cos
\varphi_j}{\zeta(e,m) - J \sum_j \cos \varphi_j} \nonumber \\ &=:& - P(\zeta(e,m))
\end{eqnarray} and
\begin{equation}
    \label{gl:sat2}
    \beta(e,m)\left[1-m^2\right] = \frac{1}{2} \int
\frac{\mb{d}^d\varphi}{(2 \pi)^d} \frac{1}{\zeta(e,m) - J \sum_j
\cos \varphi_j} =: R(\zeta(e,m)).
\end{equation}

In the following the system for a fixed energy $e$ is considered for
the limit of vanishing (positive) magnetic field $h \to 0+$, i.\,e. for
the zero-field macrostate $(e, m_{\sts\mb{sp}}(e))$.
The corresponding zero-field inverse temperature is defined to be $\beta_0(e) = \beta (e, m_{\sts
\mb{sp}}(e))$, analogously $\zeta_0(e) = \zeta (e, m_{\sts
\mb{sp}}(e))$. In the limit $m \to m_{\sts
\mb{sp}}(e)$ (corresponding to  $h(e,m) \to 0+$ for fixed energy $e$)  one gets the equation
\begin{equation}
\label{gl:spontanebed}
m_{\sts \mb{sp}}(e) (\zeta_0(e) - d J ) = 0
\end{equation}
which has to be obeyed by the spontaneous magnetisation. This
equation has the trivial solution $m_{\sts \mb{sp}}(e) = 0$. A
non-zero spontaneous magnetisation is only possible if the bracket
in (\ref{gl:spontanebed}) vanishes, i.\,e. if $\zeta_0 = dJ$. 
Therefore, an non-zero spontaneous magnetisation below some critical
energy  $e_{\sts \mb{c}}$ is only possible if $\zeta_0(e) = dJ$ for
$e<e_{\sts \mb{c}}$.

The possible emergence of a non-zero spontaneous
magnetisation is further discussed in subsection
\ref{kap:spon}. Before this consideration can be carried out it has
to be investigated whether it is possible at all to have a critical
energy $e_{\sts \mb{c}}$ larger than the ground state energy $e_{\sts
  \mb{g}} = -dJ$ so that $\zeta_0(e) = dJ$ holds below $e_{\sts \mb{c}}$.  
To this end consider first the second relation (\ref{gl:sat2}) and assume that
there exists a critical energy $e_{\sts \mb{c}}$ above which the
spontaneous magnetisation $m_{\sts \mb{sp}}(e)$ vanishes. Then one
must have  $\zeta_0(e) > dJ$ for $e>e_{\sts \mb{c}}$ and the
associated inverse temperature $\beta_0(e) = \beta(e,m_{\sts \mb{sp}}(e)=0
)$  of the zero-field macrostates $(e>e_{\sts \mb{c}},0)$ is given by
$\beta_0(e) = R(\zeta(e,0))$ (see relation (\ref{gl:sat2}) for the
definition of $R$). 
In view of equation (\ref{gl:spontanebed}) a non-zero spontaneous
magnetisation can appear for energies below the critical value only if
$\zeta_0(e) = dJ$ for all $e < e_{\sts \mb{c}}$. At the critical
energy $e_{\sts \mb{c}}$ one therefore has $\zeta (e_{\sts \mb{c}}) = \zeta_{\sts \mb{c}}$  and
the associated critical inverse temperature $\beta_{\sts \mb{c}}$ is thus
given by $\beta_{\sts \mb{c}} = R(\zeta_{\sts \mb{c}})$.
In the limit $\varphi_j \to 0$, $j=1,\ldots,d$, the sum
$\sum_j\cos{\varphi_j}$ tends to $d$ and thus the vanishing
denominator in (\ref{gl:sat2}) might cause a diverging integral for
the limit $\zeta_0 \to \zeta_{\sts \mb{c}}$. This would have the
consequence that the critical inverse temperature $\beta_{\sts \mb{c}}$ is
infinite so that the model has not phase transition.
To investigate this situation more explicitly consider the
contributions to the integral which arise from small $\varphi_j$.
In the regime $\varphi_j \to 0$ the denominator can be approximated by
$\frac{1}{2}(\varphi_1^2+\ldots+\varphi_d^2)$. Introducing polar
coordinates for the $d$-dimensional $\varphi$-space 
and excluding a small sphere of radius $\delta$ from the
integration over $\varphi_j$, $j=1, \ldots,d$, one gets the factor
\begin{equation}
\label{gl:factordiv} -\lim_{\delta\to 0 } \int_\delta
\mb{d}\varphi  \varphi^{d-3}
\end{equation}
to whom  the dominant small $\varphi_j$ contributions to the
integral $R$ are proportional. The modulus of the vector
$(\varphi_1, \ldots, \varphi_d)$ is here denoted by $\varphi$. For
one and two dimensions the limit (\ref{gl:factordiv}) diverges and
one has $\beta_{\sts \mb{c}} = \infty$ and the
spherical model does not undergo a phase transition in these
dimensions. From equation (\ref{gl:sat1}) it is evident that the
associated critical energy $e_{\sts \mb{c}} = -dJ$ is the ground
state energy of the spherical model. For dimensions larger than
two the limit (\ref{gl:factordiv}) exists and hence the (critical)
inverse temperature $\beta_{\sts \mb{c}}$ is finite and the model will
have a phase transition at some energy $e_{\sts \mb{c}} > -dJ$
(see subsequent subsections).

\subsection{Spontaneous magnetisation}
\label{kap:spon}

\begin{figure}
\begin{center}
\epsfig{file=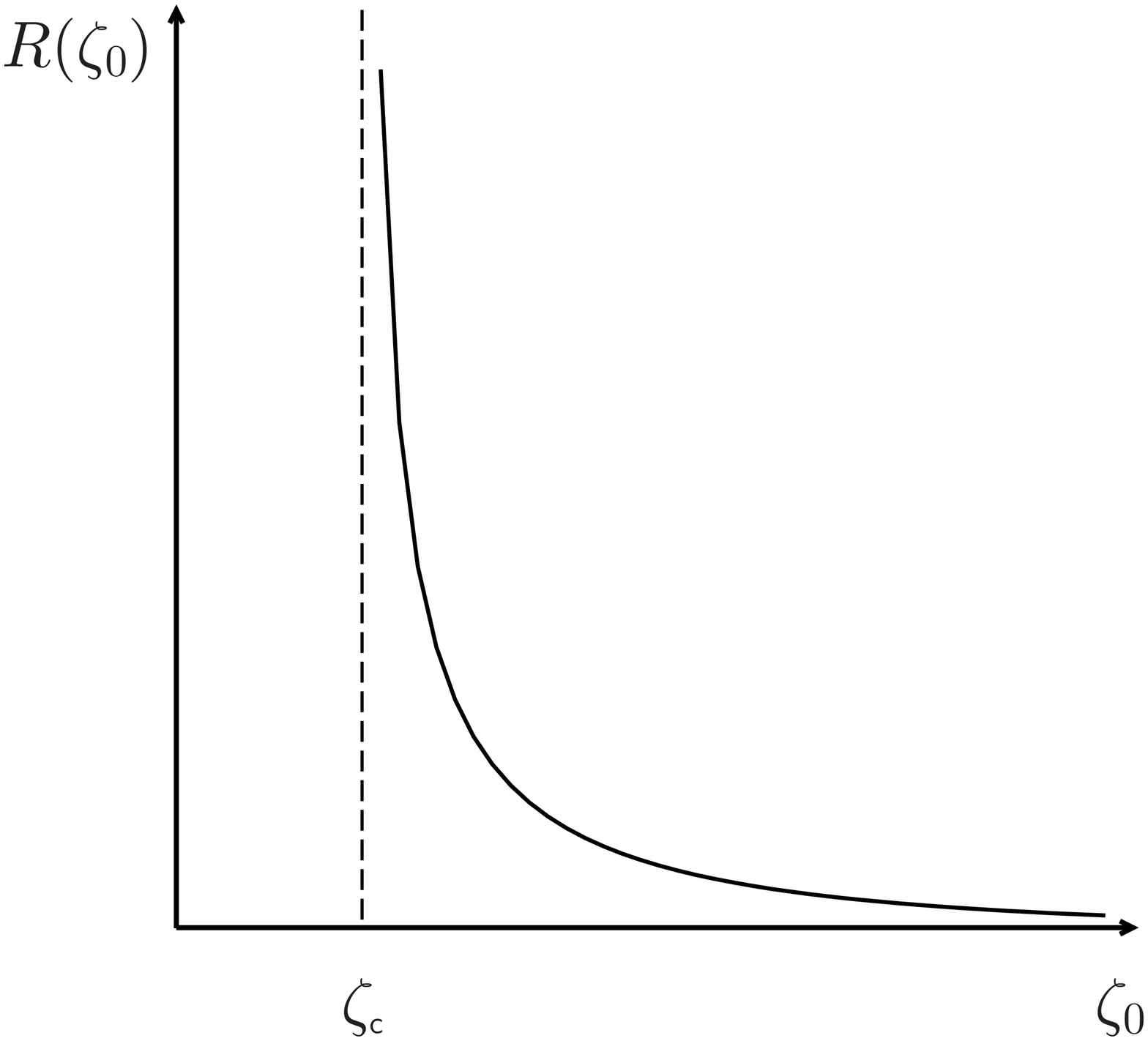,width=12em, angle=270}
\epsfig{file=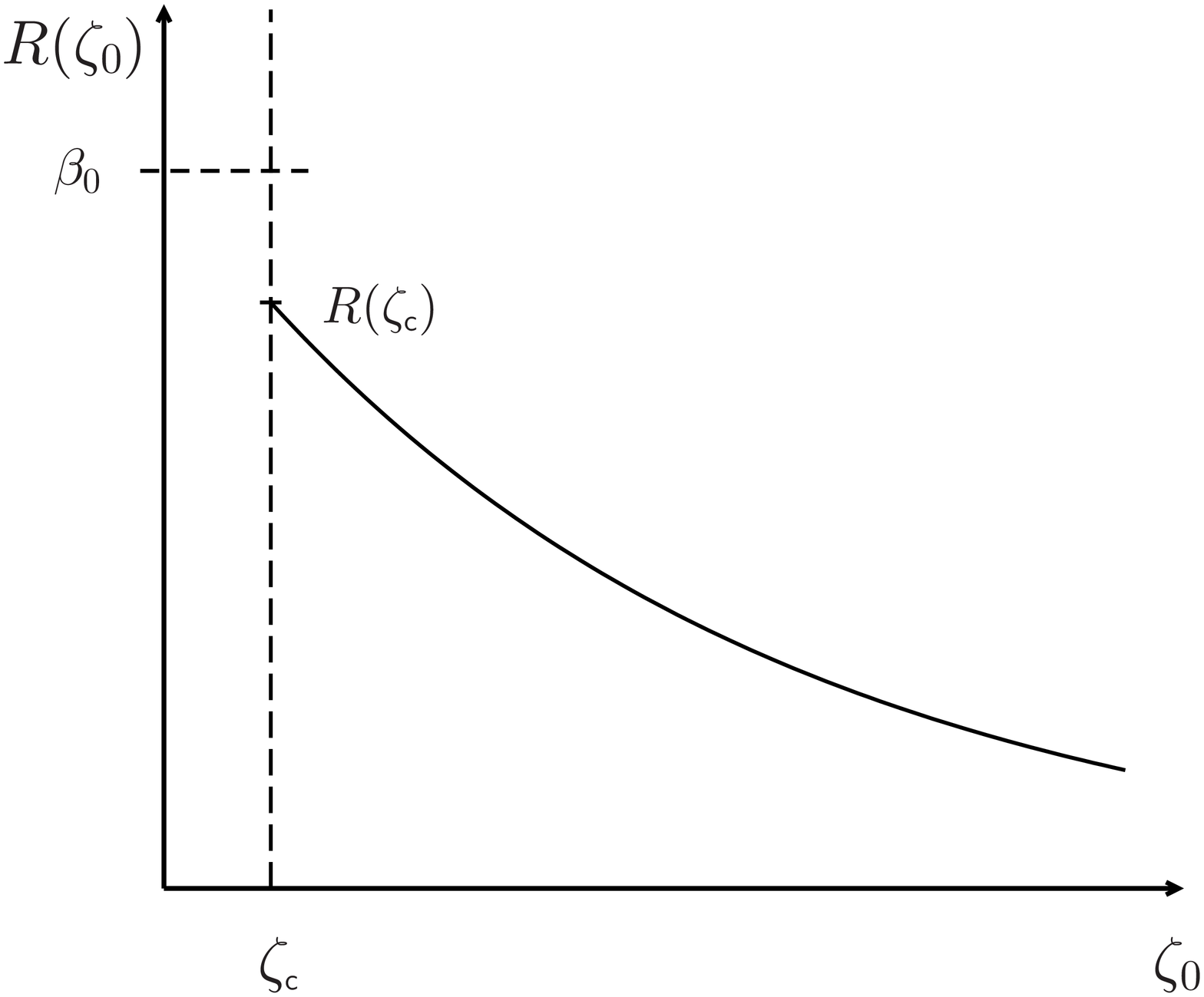,width=12em, angle=270}
\caption{\label{bild:r_schema} \small 
Schematic depiction of the saddle point equation (\ref{gl:sat2}) for
the zero-field macrostate. Dimensions $d \leq 2$ are illustrated on the left
hand side whereas the right hand side shows the situation for 
dimensions $d>2$. }
\end{center}
\end{figure}

This subsection discusses how a microcanonical
spontaneous magnetisation can emerge in the spherical model for
dimensions lager than two. Consider first the case of $d \leq 2$ for
the limit $m\to m_{\sts \mb{sp}}(e)$.
The function $R$ for the zero-field macrostate is shown schematically in figure \ref{bild:r_schema}. Below the critical
value $\zeta_{\sts \mb{c}} = dJ$ the integral $R$ is not defined.
Above $dJ$ the spontaneous magnetisation that is associated with
$\zeta_0$ has to vanish in order to satisfy equation
(\ref{gl:spontanebed}). On approaching $dJ$ from above
$R(\zeta_0)$ diverges and therefore one always has $\beta_0 =
R(\zeta_0)$ (compare equation (\ref{gl:sat2})) and no spontaneous magnetisation can emerge. The
situation for $d>2$ is also displayed in figure \ref{bild:r_schema}. In contrast
to the case $d\leq 2$ the integral $R$ approaches a finite value
$R(\zeta_{\sts \mb{c}})$ in the limit $\zeta_0 \to dJ$. If the
inverse temperature $\beta_0$ is chosen to be above $R(\zeta_{\sts \mb{c}}) =
\beta_{\sts \mb{c}}$ (or equivalently $e < e_{\sts \mb{c}}$) the
saddle point equation (\ref{gl:sat2}) can only be satisfied if a
non-zero spontaneous magnetisation $m_{\sts \mb{sp} }(e)$ shows
up.\footnote{In the microcanonical
  ensemble the energy $e$ is the natural variable and the associated
  inverse temperature $\beta_0(e)$ for the zero-field macrostate is determined
by the integral equations (\ref{gl:sat1}) and (\ref{gl:sat2}). As the
specific heat in the infinite system has a well defined sign, however,
the
inverse temperature $\beta_0$ can be chosen first and from the saddle point equations the
associated energy $e$ can be calculated in principle, finally yielding the desired
function $\beta_0(e)$. } This is possible if $\zeta_0(e) = dJ$ for $e<e_{\sts \mb{c}}$
so that the subsidiary condition (\ref{gl:spontanebed}) holds.
Saddle point equation (\ref{gl:sat2}) now reduces to
\begin{equation}
\label{gl:msponnochmitT}
    m_{\sts \mb{sp}}(e) = \sqrt{\frac{\beta_0(e) - \beta_{\sts \mb{c}}}{\beta_0(e)}}.
\end{equation}
The critical energy on the other hand is given by $e_{\sts \mb{c}}
= P(\zeta_{\sts \mb{c}})/\beta_{\sts \mb{c}} $ (see relation
(\ref{gl:sat1}) for the definition of $P$) and below $e_{\sts
\mb{c}}$ the saddle point equation (\ref{gl:sat1}) for the zero-field
macrostate is just $\beta_0(e+dJm_{\sts \mb{sp}}^2) = e_{\sts
\mb{c}}\beta_{\sts \mb{c}}$. This equation can now be used to
eliminate $\beta_0$ from (\ref{gl:msponnochmitT}) and one has
\begin{equation}
\label{gl:resultatfmsp}
    m_{\sts \mb{sp}}(e) = \sqrt{\frac{e_{\sts \mb{c}}-e}{e_{\sts
    \mb{c}}+dJ}}.
\end{equation}
The spontaneous magnetisation is therefore characterised by the
critical exponent $\beta_\varepsilon = 1/2$ for all dimensions
$d>2$. Note, however, that relation (\ref{gl:resultatfmsp}) is
valid for all energies below the critical one. At the ground state
energy $e_{\sts \mb{g}} = -dJ$ the spontaneous magnetisation is one
as expected.

At this stage it should be remarked that the entropy $s(e,m)$ does 
exist for the spontaneous magnetisation $m_{\sts \mb{sp}}(e)$ although the
argument of the logarithm in (\ref{gl:phivorsattelaus}) then
vanishes. The investigation of physical
properties corresponding to  derivatives of the entropy function
gives sensible results for the limit $m \to m_{\sts \mb{sp}}(e)$. Note
that similarly the free energy of the spherical model in the canonical formalism
is also defined for zero magnetic field (i.\,e. for the spontaneous
magnetisation) for temperatures below the critical one (e.\,g. \cite{Thompson_1988, Yan_Wannier_1965}).   

\subsection{Specific heat for dimensions $d>2$}

The two saddle point equations (\ref{gl:sat1}) and (\ref{gl:sat2})
for the  zero-field macrostate contain the spontaneous magnetisation which can be
zero or non-zero depending on whether the energy is above or below
the critical value $e_{\sts \mb{c}}$. Equation (\ref{gl:sat2}) can
be used to eliminate the magnetisation from relation
(\ref{gl:sat1}) yielding
\begin{equation}
\label{gl:eohnemGl}
    e = \frac{1}{2} \int
\frac{\mb{d}^d\varphi}{(2 \pi)^d} \frac{dJ-J\sum_j \cos
\varphi_j}{\zeta_0(e) - J \sum_j \cos \varphi_j}
\frac{1}{\beta_0(e)}- dJ.
\end{equation}
Below $e_{\sts \mb{c}}$ one has $\zeta_0 = dJ$ so
that
\begin{equation}
    e = \frac{1}{2} \frac{1}{\beta_0(e)} - dJ = \frac{1}{2} T_0(e) -dJ
\end{equation}
where $T_0(e) = 1/\beta_0$ is the actual temperature associated
with the zero-field macrostate $(e, m_{\sts \mb{sp}}(e))$. Thus, the microcanonical specific heat is
constant, namely $c_0=1/2$, for all energies below $e_{\sts
\mb{c}}$ and all dimensions $d>2$.

For arbitrary energies the microcanonical specific heat
(\ref{def:spez}) of the spherical model is given by
\begin{eqnarray}
\label{res:spezsph} c_0(e) &=& \frac{1}{2} \int
\frac{\mb{d}^d\varphi}{(2 \pi)^d} \frac{dJ-J\sum_j \cos
\varphi_j}{\zeta_0(e) - J \sum_j \cos \varphi_j} \nonumber \\ &&+  \frac{1}{2}
\beta_0(e) \int \frac{\mb{d}^d\varphi}{(2 \pi)^d} \frac{dJ-J\sum_j
\cos \varphi_j}{(\zeta_0(e) - J \sum_j \cos \varphi_j)^2}
\frac{\mb{d}\zeta_0}{\mb{d}\beta_0}
\end{eqnarray}
for the  zero-field macrostate in view of
equation (\ref{gl:eohnemGl}). For $e < e_{\sts
\mb{c} }$ the expression (\ref{res:spezsph}) reduces to the result
already discussed above as ${\mb{d}\zeta_0}/{\mb{d}\beta_0} = 0$.
In the limit $\zeta_0 \to \zeta_{\sts \mb{c}}+$ from above the two
integrals that appear in (\ref{res:spezsph}) both approach finite
values which can be seen by a similar analysis carried out for the
investigation of the behaviour of the integral $R$ in this limit.
The behaviour of the specific heat for the regime $e \to e_{\sts
\mb{c}} +$ corresponding to $\zeta_0 \to \zeta_{\sts \mb{c}}+$ is
thus determined by the behaviour of the derivative
${\mb{d}\zeta_0}/{\mb{d}\beta_0}$.

As a first step for the analysis of the limit $e\to e_{\sts
\mb{c}}+$, i.\,e. $\varepsilon \to 0+$, define for zero-field
macrostate the new variables $\tau := \beta_{\sts \mb{c}} - \beta_0$
and $\xi := \zeta_0 - \zeta_{\sts \mb{c}}$. Then $\tau =
  R(\zeta_{\sts \mb{c}})-R(\zeta_0)$ for energies above the
critical value and one has the asymptotic relations
\begin{equation}
\label{eq:r_asympt} \tau =  R(\zeta_{\sts \mb{c}})-R(\zeta_0) \sim
\left \{
\begin{array}{ll}
\xi^{(d-2)/2} & \textrm{if $2 < d < 4$,} \\
- \xi \ln \xi & \textrm{if $d = 4$,}\\
\xi & \textrm{if $d >4$}
\end{array}
\right.
\end{equation}
for the limit $\xi \to 0+$ \cite{Joyce_1972, Baxter_1982, Thompson_1988}. The deviation of the energy from the critical energy is
given by
\begin{equation}
    \varepsilon = e - e_{\sts \mb{c}} =
    \frac{P(\zeta_0(e))}{\beta_0(e)}- \frac{P(\zeta_{\sts \mb{c}})}{\beta_{\sts
    \mb{c}}} \stackrel{|\tau| \ll 1}{\sim} \frac{\tau P(\zeta_{\sts \mb{c}})}{\beta^2_{\sts \mb{c}}}
    + \frac{1}{\beta_{\sts \mb{c}}} (P(\zeta_0(e)) - P(\zeta_{\sts \mb{c}})).
\end{equation}
The two differences $P(\zeta_0) - P(\zeta_{\sts \mb{c}})$ and $R(\zeta_{\sts \mb{c}}) - R(\zeta_0)$, however, have
the same asymptotic behaviour for $\xi \to 0+$ as the two
integrals $P$ and $R$ differ only in the enumerator which does not
alter the asymptotic behaviour for $\xi \to 0+$. Thus, one has
$\varepsilon \sim \tau$ near the critical point. Note that at this
stage it is already obvious that the critical exponents of the
spherical model are the same for both the canonical and the
microcanonical ensemble.

Using these results and the relation 
\begin{equation}
    \frac{\mb{d}\zeta_0}{\mb{d}\beta_0} =\frac{\mb{d}\xi}{\mb{d}\beta_0}= - \frac{\mb{d}\xi}{\mb{d}\tau}
\end{equation}
one obtains the following asymptotic relations for the
microcanonical specific heat near the transition point:
\begin{equation}
\label{eq:spez_asympt} c_0(e) \sim \left \{
\begin{array}{lll}
{1}/{2} - A_d\xi^{-(d-4)/2} & \sim 1/2 - B_d\varepsilon^{-\frac{d-4}{d-2}} & \textrm{if $2 < d < 4$,} \\
{1}/{2}  - A_d/|\ln \xi| & \sim 1/2 -B_d/|\ln \varepsilon| & \textrm{if $d = 4$,}\\
{1}/{2} - B_d && \textrm{if $d >4$}
\end{array}
\right.
\end{equation}
where $A_d$ and $B_d$ are some constants. In the analysis of the
regime $e \to e_{\sts \mb{c}}$ the first integral in
(\ref{res:spezsph}) has been replaced by $1/2$. Note that the
correction term originating from this integral in
(\ref{res:spezsph}) does not contribute to the leading asymptotic
behaviour of the microcanonical specific heat. For dimensions
$2<d\leq 4$ the form (\ref{eq:spez_asympt}) of the microcanonical
specific heat is characterised by the negative critical exponent
$\alpha_\varepsilon = (d-4)/(d-2)$ corresponding to a
(right-sided) cusp singularity. For dimensions $d>4$ the
microcanonical specific heat has a discontinuity at the
transition point.

\subsection{Susceptibility}

The susceptibility is generally defined  by
\begin{equation}
    \chi = \frac{\mb{d}m}{\mb{d}h} =
    \left(\frac{\mb{d}h}{\mb{d}m}\right)^{-1}.
\end{equation}
Focusing on energies $e > e_{\sts \mb{c}}$ and using the general
relation (\ref{gL.hgleichungmitm}) between the magnetic field and
the magnetisation one gets the microcanonical susceptibility
\begin{equation}
    \chi_0(e) = \frac{1}{2(\zeta_0(e) - dJ)} = \frac{1}{2\xi}
\end{equation}
for the zero-field macrostate.
Expressed in terms of the energy deviation $\varepsilon$ one gets
thus the asymptotic relations
\begin{equation}
\label{eq:susz_asympt} \chi_0(e) \sim \left \{
\begin{array}{ll}
\varepsilon^{-\frac{2}{d-2}} & \textrm{if $2 < d < 4$,} \\
\left( \varepsilon /|\ln \varepsilon|\right)^{-1} & \textrm{if $d = 4$,}\\
\varepsilon^{-1} & \textrm{if $d >4$}
\end{array}
\right.
\end{equation}
for the microcanonical susceptibility in the regime $\varepsilon
\to 0+$. The microcanonical exponent of the susceptibility in $2<
d < 4$ space dimensions is therefore $\gamma_\varepsilon =
2/(d-2)$ and in dimensions $d\geq 4$ one has the microcanonical
exponent $\gamma_\varepsilon = 1$.

\subsection{Critical field}

At the critical energy $e_{\sts \mb{c}}$ the saddle point equation
(\ref{gl:sat2}) reduces to
\begin{equation}
    \beta(e_{\sts \mb{c}}, m) = R(\zeta_{\sts \mb{c}} + h(e_{\sts
    \mb{c}},m)/(2m))
\end{equation}
where relation (\ref{gL.hgleichungmitm}) has been used to
eliminate $\zeta$ from equation (\ref{gl:sat2}). Carrying out the
asymptotic analysis of $R$ for the limit $m \to m_{\sts
\mb{sp}}(e_{\sts \mb{c}}) = 0$ one gets the values
\begin{equation}
\label{eq:delta_werte} \delta_\varepsilon = \left \{
\begin{array}{cl}
\frac{d+2}{d-2} & \textrm{if $2 < d < 4$,} \\
 3& \textrm{if $d \geq 4$}
\end{array}
\right.
\end{equation}
for the critical field exponent
$\delta_\varepsilon$.\footnote{Note that the analysis of the
exponent $\delta_\varepsilon$ is similar to the canonical analysis
of $\delta_t$ (see e.\,g. \cite{Thompson_1988}) and is therefore not
displayed here.}

\section{Mean spherical model}
\label{kap:mean}

The calculations presented in the above sections showed that the
thermodynamic properties of the spherical model are equivalent in the
canonical and microcanonical ensemble as far as the spontaneous
magnetisation, the susceptibility, the specific heat and the critical
field are considered. This is apparent
at first sight from the corresponding expressions for these
quantities.  
In the microcanonical ensemble
the energy and the magnetisation are fixed to some values. These
constraints are represented by the first two delta functions in
the expression (\ref{gl:entro_deltarestriction}) for the density of
states. These two constraints can be relaxed so that they are only satisfied
on the average. This  corresponds to the canonical treatment where the Lagrange
parameters which are then introduced to satisfy the constraints on the
energy and magnetisation turn out to be related to the temperature and the
magnetic field, respectively. In a similar way the spherical constraint
(\ref{gl:nebenbedsph}) can be relaxed so that it is only satisfied on
the average (the model is then often called mean spherical
model). Lewis and Wannier used this relaxation to calculate
the properties of the canonical spherical
model \cite{Lewis_Wannier_1952}. 
However, it turned out that the properties of the mean
spherical model are only partially equivalent to the properties of the
spherical model with the rigid constraint (\ref{gl:nebenbedsph}) as
quantities which are related to the fluctuations in
the spherical constraint are different for the spherical and mean
spherical model \cite{Lewis_Wannier_1953, Lax_1955, Yan_Wannier_1965,
  Kac_Thompson_1977}. 
These differences have their origin in the fact that
the Lagrange parameter that controls the constraint in the mean
spherical model does not have a corresponding variable in the spherical
model. Note that this is different for the Lagrange parameters
temperature and magnetic field in the canonical ensemble which have
corresponding variables, namely the energy and the magnetisation.

A similar treatment can be carried out in the microcanonical
ensemble. The spherical constraint can be relaxed and controlled by a
Lagrange parameter so that it is satisfied on the average. Only two
delta functions, namely those which fix the energy and the
magnetisation, are left in the expression for the density of states
and therefore only relations (\ref{gl:z-sattelgl}) and (\ref{gl:w-sattelgl}) show up
as saddle point equations in the thermodynamic limit. However, the
requirement that the spherical constraint is satisfied on the average
leads to a subsidiary condition which fixes the introduced Lagrange
parameter. This additional subsidiary condition is identical to relation
(\ref{gl:u-sattelgl}) and therefore the investigated properties of the microcanonical
spherical model and the microcanonical mean spherical model are
equivalent. Nevertheless, the relaxation of the spherical constraint
now allows for fluctuations in this constraint and having the canonical
results in mind it might be expected
that differences occur when comparing quantities that are related to
those fluctuations. This interesting question goes beyond the scope of
the present work and is left to future studies.

\section{Summary} \label{kap:zus}

The various statistical ensembles are equivalent for systems with
short-range interactions undergoing a continuous phase transition in
the infinite volume limit. Therefore, physical quantities can be
calculated in different ensembles leading to the same thermodynamic properties. The properties of a model
system undergoing a continuous phase transition in the
thermodynamic limit can hence be directly deduced from the
microcanonical entropy. As an example the spherical model was
investigated microcanonically in this work to demonstrate these
general properties and to provide an analytic treatment of a model
system within the microcanonical formalism. The calculation of the
microcanonical entropy as presented in section \ref{kap:spezentro}
somehow resembles the determination of the free energy in the
canonical treatment (see e.\,g. \cite{Baxter_1982}). This already hints at
the equivalence of the microcanonical and the canonical
treatment of the spherical model. However, in the microcanonial calculation further saddle
point equations emerge due to the restrictions on the energy and the
magnetisation represented by the delta functions in
(\ref{gl:entro_deltarestriction}). To establish ensemble equivalence
for the spherical model explicitly these additional saddle point
equations have to be analysed as done in section
\ref{kap:kriteigen}.

The natural variables of the canonical and microcanonical
formalism are different so that one gets in general different critical
exponents for physical quantities. However, the canonical and
microcanonical critical exponents are related to each other to
ensure ensemble equivalence. In the case of a cusp singularity in
the specific heat at the transition point the microcanonical and
the canonical critical exponents are identical whereas the
exponents are different for systems with an algebraically
diverging specific  heat (compare relation (\ref{gl:exporel})).
The spherical model is a system with such a cusp singularity in
the specific heat. The microcanonical critical exponents are
therefore expected to be identical to the canonical critical
exponents. This was indeed verified explicitly in this work.

The microcanonical treatment of the spherical model on a infinite
hyper-cubic lattice carried out in this work also
demonstrates how properties of physical systems near a continuous phase
transition point can be deduced directly from the density of
states (or equivalently the microcanonical entropy) without going
to the canonical ensemble.

\ack

The author wants to thank Alfred H\"uller and Michel Pleimling for encouraging discussions
and helpful comments on the manuscript.

\end{document}